\documentclass[twocolumn,english,showpacs,prb,floatfix]{revtex4}%
\usepackage{amsfonts}
\usepackage{amsmath}
\usepackage{amssymb}
\usepackage{graphicx}%
\begin{document}
\title{First-Principles calculation of atomic hydrogen adsorption on Be(10\={1}0)
thin films }
\author{Ping Zhang, Hong-Zhou Song}
\affiliation{Institute of Applied Physics and Computational Mathematics, P.O. Box 8009,
Beijing 100088, P.R. China}

\begin{abstract}
We present a first-principles study of the atomic hydrogen adsorption onto the
Be(10\={1}0) thin film. There are two types of Be(10\={1}0) surfaces according
to the interlayer spacing between the surface and its nearest-neighbor layer.
We show that the H adsorption features on these two kinds of surfaces are
remarkably different. The work function, averaged electrostatic potential, and
the local charge density consistently show that the charge is transferred from
H to Be for L-type (see the text below) surfaces, while the transfer process
is inverted for S-type surfaces.

\end{abstract}
\maketitle

Beryllium belongs to the $sp$-bonded \textquotedblleft simple
metasl\textquotedblright, however, it is far from nearly-free-electron like
and has many anomalous properties. For example, compared to its close
neighbor, Mg, the $c/a$ ratio for hcp Be has a very nonideal value of $1.57$
(1.62 for Mg), showing the directional bonding in Be. The bonding energy of Be
is surprisingly strong with cohesive energy 3.32 eV/atom. Also, the Debye
temperature is very high, with a value of 1440 K (Mg: 400 K), and thus quantum
effects should be more pronounced for Be in determining its zero-temperature
structural properties\cite{Laz}. The electronic structure of Be is very
profound by the following unique properties: the bulk Be shows a very low
density of states (DOS) at the Fermi energy ($E_{F}$) and thus exhibits some
degree of covalent-like bonding\cite{Pap,Yu}. However, the Be(0001) and
Be(10\={1}0) surfaces have been shown, both experimentally and theoretically,
to be a nearly-free-electron system\cite{Chu}. This large metallic enhancement
in the DOS originates from the surface states\cite{Bal1}, which accounts for
roughly 80\% of the local DOS at $E_{F}$ in the outermost layer. This
peculiarly large surface-to-bulk ratio of the local DOS (LDOS) at $E_{F}$ has
been shown to provides the basic framework to understand many observed surface
properties which deviate substantially from the bulk, including abnormally
large surface core level shifts\cite{Fei1,Ald,Joh1,Liz,Cho,Run,Glan}, giant
surface Friedel oscillations\cite{Joh,Spr,Hof1,Bri}, large surface
expansion\cite{Laz1,Vob}, and enhancement of electron-phonon
interaction\cite{Heng,Bal,Tang,Ism,Eig,Shi}.

In the present study we present a detailed first-principles study of the
adsorption properties of atomic hydrogen onto Be(10\={1}0) thin films. The
interaction of H with Be(0001) surface has been theoretically studied by
Feibelman and Stumpf\cite{Stu1} who showed that H prefers ordinary hcp
threefold sites at low coverages, while at high coverage with long-range
ordering inevitably taken into account, the H adsorption has been shown to
lead Be(0001) surface to $(\sqrt{3}\times\sqrt{3})R30^{\circ}$
reconstruction\cite{Poh}. To the best of our knowledge, the adsorption of H on
Be(10\={1}0) surface has not been studied before. Considering that the clean
Be(10\={1}0) surface, and many novel effects including surface core level
shifts, electron-phonon interaction, and large surface expansion have been
extensively revealed, we feel it necessary to give a detailed study of
adsorption properties of H adsorption onto Be(10\={1}0) surface. Also, it is
an interesting task to investigate the effect of H adsorption on the various
kinds of enhancement phenomena in Be surfaces and we will leave it for future study.

The calculations were carried out using the Vienna \textit{ab initio}
simulation package\cite{Vasp} based on density-functional theory with
ultrasoft pseudopotentials\cite{Vand} and plane waves. In the present film
calculations, free-standing Be(10\={1}0) films in periodic slab geometries
were employed. The periodic slabs are separated by a vacuum region equal to 20
\AA . In all the calculations below, a surface ($1\times1$) was employed for
the supercell slab. The Brillouin-zone integration was performed using
Monkhorse-Pack scheme\cite{Pack} with a $11\times11\times1$ $k$-point grid,
and the plane-wave energy cutoff was set $300$ eV. Furthermore, the
generalized gradient approximation (GGA) with PW-91 exchange-correlation
potential has been employed with all atomic configurations fully relaxed. A
Fermi broadening of 0.05 eV was chosen to smear the occupation of the bands
around $E_{F}$ by a finite-temperature Fermi function and extrapolating to
$T=0$ K. Optimized atomic geometries are achieved when the forces on all the
unconstrained atoms are in magnitude smaller than 0.01 eV/\AA . First the
total energy of the bulk hcp Be was calculated to obtain the bulk lattice
constants. The calculated $a$- and $c$-lattice parameters are $2.272$ \AA and
$3.544$ \AA , comparable well with experimental\cite{Amo} values of $2.285$
\AA and 3.585 \AA , respectively.%

\begin{figure}[tbp]
\begin{center}
\includegraphics[width=1.0\linewidth]{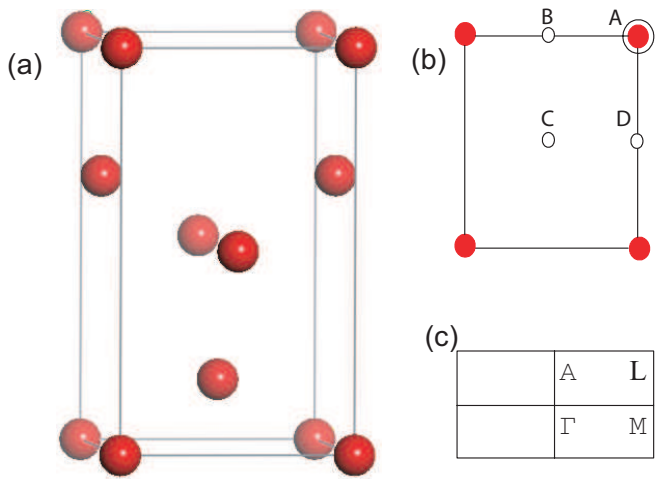}
\end{center}
\caption{(Color online) (a) Bulk unit cell of hcp Be along [10\={1}%
0] direction;
(b) The four probable adsorption sites (A, B, C, and D) for atomic hydrogen on
Be(10\={1}%
0). The red circles denote the topmost surface Be atoms; (c) the Be(10\={1}%
0) surface Brillouin zone.}
\label{fig1}
\end{figure}%
Choosing the (10\={1}0) plane as the basal plane, then the unit cell of the
bulk Be, including four atoms inside, has an orthorhombic structure with the
basis vectors given by $\mathbf{a}_{1}=(c,0,0)$, $\mathbf{a}_{2}=(0,a,0)$,
$\mathbf{a}_{3}=(0,0,\sqrt{3}a)$. In terms of this basis set, the positions of
four atoms in bulk hcp Be are given by $(0,0,0)$, $(0.5,0.5,1/6)$,
$(0,0.5,0.5)$, and $(0.5,0,2/3)$. One can see that the (10\={1}0) surface of
hcp Be can be terminated either with a short first interlayer spacing,
$d_{s}=\sqrt{3}a/6$, or with a long one, $d_{l}=\sqrt{3}a/3$. The preferred
termination has been found experimentally and theoretically to be the short
one with $d_{s}$\cite{Hof}. During the following discussions, we will call a
Be(10\={1}0) monolayer as a S layer if the interlayer spacing between this
layer and its nearest-neighbor layer (from below) is $d_{s}$. In the same way,
a Be(10\={1}0) monolayer will be termed in this paper as a L layer if the
interlayer spacing between this layer and its nearest-neighbor layer (from
below) is $d_{l}$.%

\begin{figure}[tbp]
\begin{center}
\includegraphics[width=1.0\linewidth]{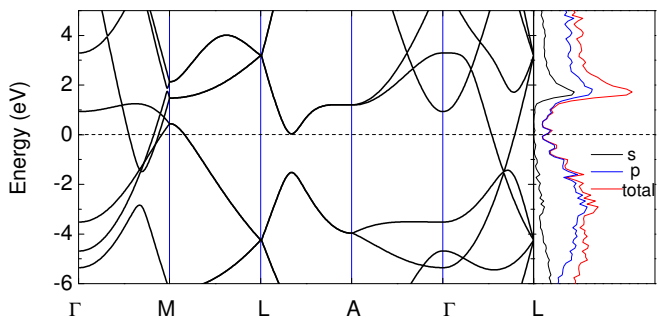}
\end{center}
\caption
{(Color online) GGA energy bands and density of electron states (right panel) of hcp bulk Be.
The dashed line denotes Fermi level.} \label{fig2}
\end{figure}%
The band structure and the DOS of bulk Be with unit cell in Fig. 1(a) are
shown in Fig. 2. One can see that the density of electronic states for bulk Be
resembles somewhat that of a semiconductor since it has a minimum at the Fermi
energy. This makes Be different from its close neighbors, such as Mg, whose
band property is nearly free-electron like. The bulk Be bands display a direct
gap in a large range of the Brillouin zone, unlike Mg. Mg has a filled state
at $\Gamma$ with energy $\simeq1.3$ eV, while the corresponding Be state is
above the Fermi energy and its band is nearly flat. This band is the source of
both the low density of states near the Fermi energy and the high peak above
the Fermi energy\cite{Laz}. Although the electronic configuration of elemental
Be is $1s^{2}2s^{2}$, one can see from Fig. 2 that the $p$-orbital component
in bulk Be plays a main role around $E_{F}$. The bonding properties of hcp Be
is anisotropic, which can be seen by the fact that the $c/a$ ratio (1.56) is
one of the most contracted for hcp metals (for Mg, $c/a\simeq1.62$). Thus
out-of-plane neighbors have shorter bonds than in-plane neighbors. Another
evidence of this anisotropy is that the contribution of $p_{x}$ and $p_{y}$
orbitals to the DOS is different from that of $p_{z}$ component\cite{Cohen}.%

\begin{figure}[tbp]
\begin{center}
\includegraphics[width=1.0\linewidth]{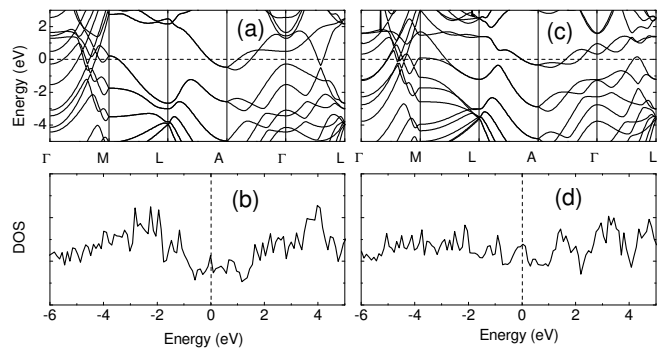}
\end{center}
\caption{Energy bands and density of electron states of Be(10\={1}%
0) thin films with 10
monolayers included in the slabs. The left panels denote the case that the bottom and top surfaces
of the slab are L-type, while the right panels denote the case of S-type surfaces.}
\label{fig3}
\end{figure}%
The above-mentioned \textquotedblleft semiconducting\textquotedblright%
\ metallic picture for bulk Be is prominently changed in the case of
Be(10\={1}0) surfaces. To illustrate this change, the electronic structure
properties of Be(10\={1}0) thin films with two kinds of surfaces are shown in
Fig. 3, wherein the left panels give the band structure and density of
electron states for a structure with L-type surfaces, while the right panels
show the results for a structure with L-type surfaces. The number of
Be(10\={1}0) monolayers in these two slabs has been fixed to be 10. Compared
to the bulk results shown in Fig. 2, it reveals in Fig. 3 that (i) the sharp
DOS peak above Fermi level in Fig. 2 becomes broad for Be(10\={1}0) thin
films, resulting in a prominent enhancement of the DOS at $E_{F}$ with respect
to the bulk case. This phenomenon of high density of surface states at $E_{F}$
was previously reported for Be(0001)\cite{Chu}, and was also noticed in other
metal films such as W(110) and Mo(110)\cite{Rot}; (ii) Compared to the L-type
surfaces, the enhancement of DOS at $E_{F}$ is more prominent for S-type
surfaces, implying that the metallic behavior in S-type Be(10\={1}0) surfaces
is more strong than in L-type surfaces. Also, the band structure for S-type
surfaces deviates more far from the bulk band structure than that for L-type
surfaces. Thus the preferrence of S-type surfaces to L-type surfaces, as
observed in experiment\cite{Hof}, is the result of this more metallic behavior
of S-type surfaces; (iii) The band structure aroung $\Gamma$ point is
characterized by a series of subbands which can be well fit by $E_{n}%
+\hbar^{2}(k_{x}^{2}+k_{y}^{2})/2m^{\ast}$ with the effective mass $m^{\ast}$,
demonstrating from another aspect the quasi-2D free-electron character in the
Be(10\={1}0) thin film, which is contrary to the bulk Be.

Now we focus our attention to the atomic hydrogen adsorption onto
Be(10\={1}0). Before we study the adsorption properties as a function of the
thickness of Be(10\={1}0) thin films, we need to determine the energetically
favourable adsorption site. Since the preference of adsorption site is not
sensitive to the thickness of the substrate, thus to look for this preference,
it is sufficient to give a study on the slabs with fixed thickness of the
Be(10\={1}0) film. As mentioned above, there are two kinds of Be(10\={1}0)
surfaces, namely the S- and L-type surfaces. For both of these two kinds of
surfaces the thickness of the Be(10\={1}0) film is fixed to be 9 monolayers.
We choose four most probable adsorption sites, which have been indicated in
Fig. 1(b). The binding energy is calculated using the following equation:
Binding energy [atomic H]$=-$($E[$H/Be(10\={1}0)$]-E[$Be$(10\bar{1}%
0)]-2E[$H$]$)/2 where $E[$H/Be(10\={1}0)$]$ is the total energy of a slab
which consists of 9 layers of Be atoms and one H atom on each side,
$E[$Be$(10\bar{1}0)]$ is total energy of the slab without H atoms, and
$E[$H$]$ is total energy of a free H atom which is put in a 10 \AA $\times$10
\AA $\times$10 \AA supercell. As a result, the calculated binding energy of
atomic H for these four different adsorption configurations in Fig. 1(b) is
2.38 eV (A), 3.51 eV (B), 3.42 eV (C), 2.44 eV (D) for the L-type surface, and
1.44 eV (A), 2.2 eV (B), 1.61 eV (C), 2.06 eV (D) for the S-type surface,
showing a clear preference for B site adsorption for both kinds of surfaces.
At this site the H atoms interact strongly with the dangling bond like $p$
orbitals of the Be surface atoms. Those orbitals point out of the surface.%

\begin{figure}[tbp]
\begin{center}
\includegraphics[width=1.0\linewidth]{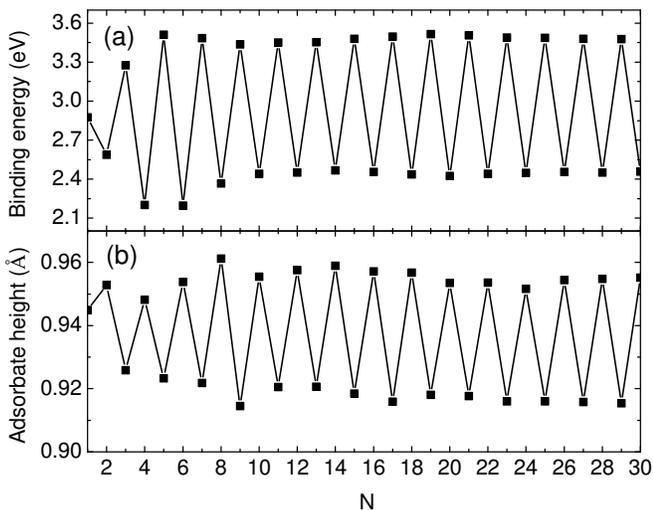}
\end{center}
\caption
{Calculated (a) binding energy of H adatom and (b) adsorbate height as a function the number of Be(10\={1}%
0) monolayers
in the slab.}
\label{fig4}
\end{figure}%
After finding the preferred atomic H adsorption site, we give a series of
calculations for the binding energy of the H adsorbate as a function of the
thickness of Be(10\={1}0) thin films. The results are summarized in Fig. 4(a).
One can see that the binding energy oscillates when increasing the Be
monolayers in the slab. The peaks in the oscillations correspond to the case
that the topmost Be(10\={1}0) monolayer is L-type, while the valley values are
for the H adsorption onto the S-type surfaces. In experiment this dependence
can be observed by investigating the dependence of H coverage on the
monolayers of Be(10\={1}0) thin films. The cohesion between the H overlayer
and the L-type Be(10\={1}0) surface is so strong that the majority of
available electrons at the interface participate in the metallic behavior with
only a minority participating in covalent H-Be bonds. For H adsorption onto
the S-type Be(10\={1}0) surface, whereas, the binding energy is not so strong,
implying that the surface bonding is dominated by the covalent hybridization
between H $s$ and Be $p$ orbitals with less ionic behavior. The height of H
adatom above the Be(10\={1}0) surface is plotted in Fig. 4(b), also showing an
oscillatory variance between the S- and L-type surfaces. The typical H
adsorbate height is 0.915 \r{A}for L-type surfaces and 0.955 \r{A}for S-type
surfaces. The lower value of adsorbate height for L-type surfaces compared to
S-type surfaces implies more attraction between adatom H and topmost Be atoms
for L-type surfaces.%

\begin{figure}[tbp]
\begin{center}
\includegraphics[width=1.0\linewidth]{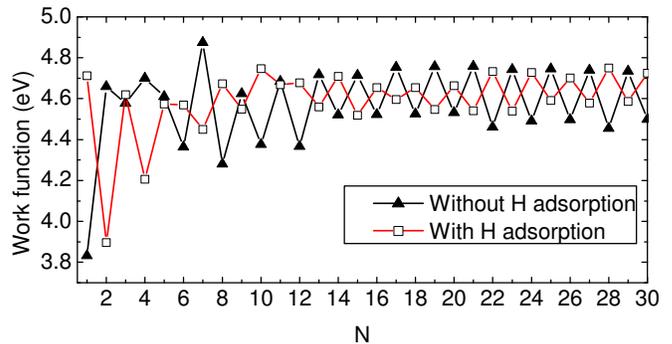}
\end{center}
\caption{(Color online) Work function of Be(10\={1}%
0) thin film with (open squares) and without (filled triangles) H adsorption as a function of the number of Be monolayers in
the slab. The bottom layer has been fixed to be L-type}
\label{fig5}
\end{figure}%
Figure 5 shows the work function of the Be(10\={1}0) film as a function of the
number of Be monolayers in the slab with and without H adsorption,
respectively. During the calculation the bottom Be layer in the slabs is fixed
to be L-type. In the absence of H adsorption, one can see that the work
function for the clean Be(10\={1}0) surfaces is oscillatory with respect to
the thickness of the film. There is a critical value for the number $N$ of
Be(10\={1}0) monolayers in the slab, separating Fig. 5 into two regimes: when
$N<6$, we find that the work function is much larger for the S-type surfaces
than for the L-type surfaces, resulting in the clear interlayer oscillations
of the work function as shown in Fig. 4. For convenience of expression, we
call this kind of oscillation as SL-type; When the value of $N$ is larger than
$6$, the work function for the Be(10\={1}0) L-type surfaces turns out to be
much larger than that for the S-type surfaces, resulting in a converted
interlayer oscillations of the work function as shown in Fig. 5. Similarly we
call this kind of oscillation as LS-type. In the presence of H adsorption, one
can see from Fig. 5 that the situation is totally inversed: in the case of
ultra-thin films ($N<6$), the work function is oscillatory in a way that its
value is larger for L-type surfaces than for S-type surfaces, while in the
case of $N>6$, the work function for L-type surfaces is much lower than that
for the S-type surfaces.

Clearly, in the absence (presence) of H adsorption, the transition from a SL
(LS) oscillation at $N<6$ to a LS (SL) oscillation at $N>6$ is due to the
atomic relaxation process of Be(10\={1}0) monolayers in the slab. In fact, in
another series of calculation, by fixing the bottom Be surface in the slabs to
be S-type, we have found that the critical value for the oscillation
transition is decreased to be $N=3$. This reduction of critical value of $N$
is not difficult to understand when one recalls that the preferred
Be(10\={1}0) surface is S-type\cite{Hof}. Thus we arrive at the conclusion
that the steady oscillation mode of Be(10\={1}0) work function in the absence
(presence) of H adsorption is LS- (SL-) type.

The change of oscillation mode of work function by the presence of H
adsorption indicates a partially ionic character of the H-Be bond, i.e.,
charge is transferred between H and Be. However, the transfer direction is
remarkably different for the two types of surfaces. It is generally believed
that the work functions increase (decrease) when the negatively (positively)
charged adatom adsorbs above the surface. For H adatom on L-type Be(10\={1}0)
surfaces, since the work function is reduced, thus charge is transferred from
H to Be, leading to a polarization of electronic charge away from the H
adlayer towards the Be surface. For H adsorption on S-type Be(10\={1}0)
surfaces, whereas, since the work function is increased, thus in this case
charge is transferred from Be to H and the H adsorbate is negatively charged.%

\begin{figure}[tbp]
\begin{center}
\includegraphics[width=1.0\linewidth]{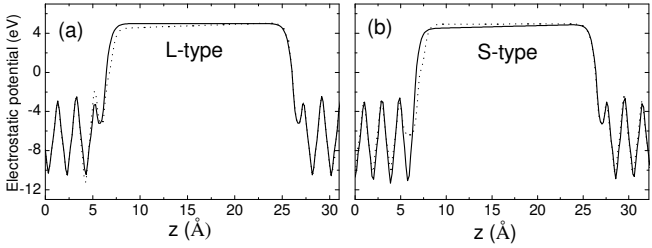}
\end{center}
\caption
{Planar-averaged electrostatic potential of clean (solid line) and H-adsorbed (dotted line)
Be(10\={1}%
0) thin films. The left and right panels are the results for S- and L-type
Be(10\={1}0) surfaces, respectively.}
\label{fig6}
\end{figure}%
To show more clearly this conversion from LS- to SL-type oscillations in work
function by the presence of H adsorption, we further plot in Fig. 6 the
$xy$-averaged electrostatic potentials $\bar{V}(z)$ as a function of $z$
coordinate normal to the Be(10\={1}0) surface for typical L-type [Fig. 6(a)]
and S-type [Fig. 6(b)] Be(10\={1}0) surfaces. The solid lines Fig. 6 are the
results for the clean surfaces while the dotted lines account for the H
adsorption. From Fig. 6(a) one can see that in the presence of H adsorption,
the electrostatic potential for L-type surface curves down towards the
surface, showing the existence of a surface dipole due to the charge transfer
from H to Be. For the S-type surface, as shown in Fig. 6(b), however, one can
see that $\bar{V}(z)$ curves up towards the surface by the H adsorption. In
this case, the surface dipole, due to the charge transfer from Be to H, points
to the opposite and increases the work function, which subsequently displays
an SL-type oscillation mode.%

\begin{figure}[tbp]
\begin{center}
\includegraphics[width=1.0\linewidth]{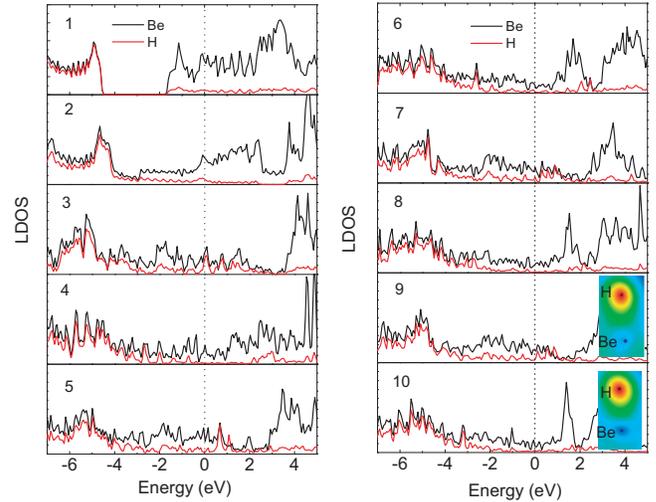}
\end{center}
\caption
{(Color online) Calculated local density of states of H adatom and topmost
Be atom. The number of Be(10\={1}%
0) monolayers in the slabs is shown in the figure. The
even numbers (2, 4, 6, 8, 10) denote the cases that the topmost Be(10\={1}%
0) monolayer
is L-type, while the odd numbers denote the cases of S-type surfaces}
\label{fig7}
\end{figure}%
Now we turn to a full study of the surface electronic structure of H-covered
Be(10\={1}0). For this we calculate the local density of states (LDOS) through
projections of the total wave function onto atoms of interest within the
Wigner-Seitz spheres around them. Figure 7 plots a series of LDOS of H adatom
layer and the topmost Be(10\={1}0) layer for different thickness of Be thin
films. The even numbers (2, 4, 6, 8, 10) in Fig. 7 denote the cases that the
topmost Be(10\={1}0) monolayer is L-type, while the odd numbers denote the
cases of S-type surfaces. One can see that the LDOS is largely different for
the two types of Be(10\={1}0) surfaces. The magnitude of LDOS at Fermi level
is higher for S-type surfaces than that for L-type surfaces, which means that
the charge transfer between H and Be is more active for S-type surfaces than
for L-type surfaces. This also can be seen in the insets in Fig. 7, which plot
the local charge densities for S- and L-type surfaces, respectively. One can
see that compared to the case of H adsorption on L-type surfaces, more
electrons populate around H adatom on S-type surfaces. Thus we arrive at the
conclusion that the H adatom on S-type surfaces behaves more negatively
charged. This is consistent with the result in Fig. 6 which shows that the
charge transfers from H to Be for L-type surfaces, while the transfer process
is inverted for S-type surfaces.

In summary, the H adsorption on Be(10\={1}0) thin films has been studied by
density-functional theory pseudopotential plane-wave calculations. The
dependence of electronic structure, binding energy, and workfunction upon the
thickness of the film has been fully investigated. We have shown that (i) the
S-type Be(10\={1}0) surface is more metallic than the L-type surface; (ii) The
H-adsorption induced charge transfer process is different for the two kinds
Be(10\={1}0) surfaces. While the charge is transferred from H to Be for L-type
surface, it is transferred from Be to H for S-type surface; (iii) In the
absence of H adsorption, the work function displays a LS-type oscillation mode
as a function of the thickness of the Be(10\={1}0) film. In the presence of H
adsorption, the oscillation mode is transformed to be SL-type, i.e., the value
of work function is high for S-type surface and low for L-type surface. This
transformation of the oscillation mode of the work function is closely related
to the charge transfer process mentioned in (ii). We expect this prominent
change in the work function and charge transfer will have a significant effect
on the other relevant physical properties\cite{Plu} in Be(10\={1}0) surface
with H adsorption.

\begin{acknowledgments}
This work was partially supported by CNSF No. 10544004 and 10604010.
\end{acknowledgments}

\end{document}